\documentclass{article}

\usepackage{mathtools, amssymb, mathrsfs,amsthm}
\usepackage{array, multirow, tabularx}
\usepackage[table] {xcolor}
\usepackage{enumerate}
\usepackage{float, caption}
\usepackage{lineno}

\newtheorem{theo}{Theorem}
\newtheorem{property}{Property}

\begin{document}

\title{Pseudo-linear regression identification based on generalized orthonormal transfer functions: Convergence conditions and bias distribution analysis}

\author  {Bernard Vau\thanks {  bernard.vau@satie.ens-cachan.fr}, Henri Bourl{\`e}s \\ \small{ SATIE, Ecole normale sup{\'e}rieure Paris-Saclay  94230 Cachan France} }

\maketitle

\begin{abstract}
In this paper we generalize three identification recursive algorithms belonging to the pseudo-linear class, by introducing a predictor established on a generalized orthonormal function basis. Contrary to the existing identification schemes that use such functions, no constraint on the model poles is imposed. Not only  this predictor parameterization offers the opportunity to relax the convergence conditions of the associated recursive schemes, but it entails a modification of the bias distribution linked to the basis poles. This result is specific to pseudo-linear regression properties, and cannot be transposed to most of prediction error method algorithms. A detailed bias distribution is provided, using the concept of equivalent prediction error, which reveals strong analogies between the three proposed schemes, corresponding to ARMAX, Output Error and a generalization of ARX models. That leads to introduce an indicator of the basis poles location effect on the bias distribution in the frequency domain. As shown by the simulations, the said basis poles play the role of tuning parameters, allowing to manage the model fit in the frequency domain, and allowing efficient identification of fast sampled or stiff discrete-time systems. 
\end{abstract}

\section{Introduction}
Algorithms dedicated to discrete-time identification are generally subdivided in three classes \cite{r11}: Prediction error methods (PEM), Instrumental variable methods (IV), and pseudo-linear regression methods (PLR). This third category presents a specific interest, especially in the perspective of recursive (on-line) identification \cite{r6}. For example, the extended recursive least-squares \cite{r14}, \cite{r15}, or the recursive output error algorithm \cite{r16}, that belong to this class are celebrated schemes that have been widely used in adaptive control \cite{r4}. A little more recently, in the nineties, several closed-loop identification structures belonging to the pseudo-linear regression class appeared \cite{r17}, \cite{r18}.  It has been emphasized that the choice of the sampling frequency is crucial in discrete-time identification \cite{r19}, and that discrete-time identification algorithms are generally not robust in fast sampling situations \cite{r20} (chap. 13). For example, some specialists of pseudo-linear regression algorithms recommend that the sampling frequency be not higher than 25-times the system expected bandwidth (for open-loop identification), and they systematically represent Bode Diagrams on a frequency linear scale \cite{r13}. Generally speaking, models obtained with PLR schemes are even less reliable than others at low frequency, which prevents from using them in a fast sampling situation. As a result, the identification of systems having modes with frequencies separated from several decades (stiff systems) is intractable with these structures. The reason for these limitations has been pointed out recently in \cite{r9}: It is a consequence of the specific PLR schemes bias distribution over frequency, which differs from the bias distribution of the corresponding PEM algorithms for a given predictor model. For example, the open-loop PLR output error and the ARMAX limit models are both weighted (for the definition of the limit model see \cite{r11}, chap. 8), exactly as the least squares algorithm limit model, for which it is well-known that the model misfit in low frequency is poorly minimized in the criterion, see \cite{r11}, pp. 268-269. In order to overcome the above limitations, this paper presents a parametrization for the regressor of the predictor developed on the orthonormal transfer function bases introduced by Heuberger et al. \cite{r3}, which are at the origin of the Hambo transform \cite{r1}, \cite{r2}. In the literature, identification schemes  using series expansion of orthonormal transfer functions (for example, in the case of Laguerre transfer function see \cite{r21}),  are made of predictors fed only by the system input \cite{r1}, thus they can be considered as a generalization of finite impulse response systems, with a specification of the model poles. Here we do not impose any poles to the estimated model, the use we make of the orthonormal transfer functions can be interpreted as -roughly speaking- a generalization of infinite impulse response systems, i.e. the predictor is fed not only with the system input, but depends also on the measured (or estimated) system output.
In the context of PLR, the parametrization proposed here has a clear impact on the convergence conditions of the identification algorithm, and the basis poles can be used as tuning parameters in order to relax the convergence of classical PLR schemes.  Furthermore, one shows that the basis poles have a crucial impact on the bias distribution -contrary to what would happen if the same parameterization were employed in the context of PEM-. The bias distribution analysis is carried out with the recently developed concept of equivalent prediction error \cite{r9}, which corresponds to the signal whose variance is effectively minimized in the PLR scheme. We demonstrate that, regarding the deterministic part, the weighting functions of the limit models are the same for the output-error, ARMAX, and a generalized version of ARX predictor. The optimization problem can be expressed in the Hambo frequency domain, in which it has always the same structure. Since this Hambo frequency scale is distorted compared with the classical frequency scale, a measure of this distortion from the logarithmic frequency scale to the Hambo frequency scale, in function of the basis poles, is provided. We show that it can be interpreted as an indicator of the basis poles effect on the bias distribution over frequency.  The simulations show that the basis poles play the role of tuning parameters, impacting the bias distribution, and making it possible to identify accurately discrete-time fast sampled or stiff systems. The identification of stiff systems is an emerging area and is reputed to be a quite challenging subject in identification, see \cite{r22}.  This paper is the first to propose a methodology dedicated to discrete-time identification of such systems.

\section{Definitions related to generalized orthonormal functions }
In this section we recall very briefly some definitions related to orthonormal transfer functions from a balanced realization of an all-pass function, as proposed in \cite{r3}. The reader interested in all theoretical aspects of these functions can refer to \cite{r1}, and \cite{r2}. Let us consider the Blashke product $G_{b}(z^{-1})$, with $G_{b}(z^{-1})G_{b}(z)=1$, such that
\begin{equation}
G_{b}(z^{-1})=\prod_{k=0}^{\eta_{p}-1}\frac{p_k-z^{-1}}{1-p_{k}z^{-1}}
\label{e001}
\end{equation}
where $p_{k}$ are the basis poles, and $\eta_p$, the poles number.\newline
This transfer function can be represented by means of a balanced state-space realization \newline 
$G_{b}(z)=D_b+C_{b}\left(zI-A_b\right)^{-1}B_b$, which satisfies
\begin{equation}
\begin{bmatrix}
A_b & B_b \\
C_b & D_b 
\end{bmatrix}^* 
\begin{bmatrix}
A_b & B_b \\
C_b & D_b 
\end{bmatrix}=I
\end{equation}
The orthonormal functions basis proposed by Heuberger and al. \cite{r3} corresponds to the vectors $V_{k}$ with size $(\eta_p,1)$, such that

\begin{subequations}
\begin{align}
V_{1}(z)=\left(zI-A_b\right)^{-1}B_b \\
V_{k}(z)=\left(zI-A_b\right)^{-1}B_{b}G_{b}^{k-1}(z) 
\label{e03}
\end{align}
\end{subequations}

\noindent These functions form a Hilbert basis of strictly proper stable transfer functions in $H_2$. The orthonormality holds because of the orthonormal state space expression of $G_{b}(z)$. 
Particular configurations of $\eta_p$ and $p_k$ correspond to well known cases: $\eta_p=1, p_{0}=0$ is the classical $z^{-1}, z^{-2}, \cdots$ basis, and \newline $\eta_p=1, \lvert p_0 \rvert <1$ corresponds to the Laguerre basis. \newline
\noindent  Consider a stable proper transfer function $H(z)$. By definition its Hambo (operator) transform denoted by $\tilde{H}(\lambda)$ is
\begin{equation}
\tilde{H}(\lambda)=\sum_{\tau=0}^{\infty}\mathcal{H}_{\tau}\lambda^{-\tau}
\end{equation}
\noindent where
\begin{equation}
\mathcal{H}_{\tau}=\oint_{\mathbb{T}}V_{1}(z)G_{b}^{\tau}(z)H(1/z)V_{1}^{T}(1/z)\frac{dz}{z}
\end{equation}
\noindent A more tractable expression is given by

\begin{equation}
\tilde{H}(\lambda)=\sum_{k=1}^{\eta_p}\frac{H(z_{k})V_{1}(z_k)V_{1}^{T}(1/z_k)}{V_1^T(z_k)V_1(1/z_k)}
\label{expr_Hambo}
\end{equation}

\noindent where $z_k$ are the eigenvalues of $A_b+B_b(\lambda-D_b)^{-1}C_b$, ($z_k$ depending on $\lambda$), $\tilde{G}(\lambda)$ being a matrix of transfer functions (expressed with $\lambda)$, of size $(\eta_p,\eta_p)$. Then the mapping from $\lambda$ to $z$ is one to $\eta_p$.\newline

\noindent Consider now a causal sequence $\{y(t)\}$ which is square summable, and define 
\[v_1(t)=A_b^{t-1}B_b \]
\[v_{k+1}(t)=G_b(q)v_k(t) \]

\noindent The Hambo signal transform of $\{y(t)\}$ is the sequence $\{\tilde{y}(k)\}$ such that
\[\tilde{y}(k)=\sum_{t=0}^{\infty}v_k(t)y(t) \]
\noindent with $\tilde{y}(k) \in \mathbb{R}^{\eta_p \times 1}$. And the $\lambda-$domain representation of the Hambo signal transform is
\[\tilde{y}(\lambda)=\sum_{k=0}^{\infty}\tilde{y}(k)\lambda^{-k} \]
\noindent This definition is similar to the definition of the $z-$transform of a signal. Let $\{u(t)\}$ be a causal square summable sequence. If
\[y(t)=G(q)u(t) \]
\noindent One has
\[ \tilde{y}(\lambda)=\tilde{G}(\lambda)\tilde{u}(\lambda) \]

\section{Optimal predictors expressed on orthonormal functions bases}
In the sequel, the following notations are employed: \newline

\noindent $\theta_0$ is the parameters vector of the true system, \newline
$\theta$ is the parameter vector of the predictor, \newline
$\hat{\theta}(t)$ is the estimated parameter vector, \newline
$\hat{\theta}^{*}$ is the limit estimated parameter vector, \newline
$\phi(t)$ is the regressor of the predictor. \newline

Let us consider $\{u(t)\}$, $\{y(t)\}$ the monovariable LTI system input and output, $\{e(t)\}$ a centered gaussian white noise, and $\{v(t)\}$ a centered noise uncorrelated with $\{u(t)\}$. According to Landau \cite{r4}, we distinguish two classes of stochastic models. The equation error model:
\begin{equation}
A(q^{-1})y(t)=B(q^{-1})u(t)+C(q^{-1})e(t) 
\label{e_ARX}
\end{equation}
where $A(q^{-1})$  is a monic polynomial in $q^{-1}$, the case $C(q^{-1})=1$ corresponding to the ARX model, and the case where $C(q^{-1})$ is a monic polynomial in $q^{-1}$, corresponding to the ARMAX model. On the other hand the output error model is given by ($v(t)$ being a disturbance uncorrelated with respect to $u(t)$) 
\begin{equation}
 A(q^{-1})y(t)=B(q^{-1})u(t)+A(q^{-1})v(t) 
\label{e_OE}
\end{equation}
Let $\hat{y}(t)$ be the predicted output, and $\varepsilon(t)=y(t)-\hat{y}(t)$, the prediction error. The optimal predictor of the equation error model is classically given by (see \cite{r4})
\begin{equation}
\hat{A}\left(q^{-1}\right)\hat{y}(t)=\hat{B}\left(q^{-1}\right)u(t)+\left(\hat{C}(q^{-1})-\hat{A}(q^{-1})\right)\varepsilon(t)
\end{equation}
where $\hat{A}\left(q^{-1}\right),\hat{B}\left(q^{-1}\right), \hat{C}\left(q^{-1}\right)$ are the estimations of polynomials \\
 $A\left(q^{-1}\right),B\left(q^{-1}\right),C\left(q^{-1}\right)$. \newline
On the other hand, the optimal predicted output of the output error model is
\begin{equation}
\hat{A}\left(q^{-1}\right)\hat{y}(t)=\hat{B}\left(q^{-1}\right)u(t)
\end{equation}

\noindent In the context of PLR identification, whatever the predictor structure is, the predicted output at time $t+1$ is written as:
\begin{equation}
\hat{y}(t+1)=\hat{\theta}^{T}(t+1)\phi\left(t,\hat{\theta}(t)\right) 
\label{e_yhat}
\end{equation}
where $\hat\theta(t)$ is the estimated parameter vector, and $\phi(t,\hat\theta(t))$, the regressor depending on past inputs and (system and/or predictor) outputs. The basic philosophy of pseudo-linear class consists in neglecting the regressor dependance with respect to $\hat\theta$ in the computation of the estimated parameter vector.
The purpose of this paper is to study identification algorithms belonging to the pseudo-linear class, when the regressor of the predictor is expressed not in function of the $\{ q^{-1},q^{-2}, \cdots \}$ basis, but on the orthonormal function basis $\{ V_1(q^{-1}), V_2(q^{-2}), \cdots \} $ basis, as defined in the previous section. That leads to consider the following expressions of the predicted output, according to the various stochastic models:
\begin{itemize}
\item Generalized-ARX predictor: 
\begin{multline}
\footnotesize{
{\hat{y}(t+1)=-\sum_{k=1}^{\frac{\eta_a}{\eta_p}}\hat{m}_k^{T}V_{k}(q^{-1})y(t+1) + \cdots}}  \\
\footnotesize{\cdots +\sum_{k=1}^{\frac{\eta_a}{\eta_p}}\hat{n}_k^{T}V_k(q^{-1})u(t+1)}
\label{e1}
\end{multline}

\item Generalized-ARMAX predictor:
\begin{multline}
\footnotesize {\hat{y}(t+1)=-\sum_{k=1}^{\frac{\eta_a}{\eta_p}}\hat{m}_k^{T}V_{k}(q^{-1})y(t+1) + \cdots}\\
\footnotesize{\dots +\sum_{k=1}^{\frac{\eta_a}{\eta_p}}\hat{n}_k^{T}V_k(q^{-1})u(t+1)+\sum_{k=1}^{\frac{\eta_a}{\eta_p}}\hat{l}_k^{T}V_k(q^{-1})\varepsilon(t+1) }  
\label{e2}
\end{multline}

\item Generalized-output error predictor:
\begin{multline}
\footnotesize{
{\hat{y}(t+1)=-\sum_{k=1}^{\frac{\eta_a}{\eta_p}}\hat{m}_k^{T}V_{k}(q^{-1})\hat{y}(t+1) + \cdots}} \\
\footnotesize{{\sum_{k=1}^{\frac{\eta_a}{\eta_p}}\hat{n}_k^{T}V_k(q^{-1})u(t+1) }}
\label{e3}
\end{multline}

\end{itemize}

\noindent where $\eta_a$ is the predictor order, and we assume that it is a multiple of $\eta_p$, $\hat{m}_k, \hat{n}_k, \hat{l}_k$ the estimated parameter vector (size $(\eta_p,1)$).  As the orthonormal transfer function vectors $V_k(q^{-1})$ are strictly proper, there is no algebraical loop in expressions \eqref{e1}, \eqref{e2}, and \eqref{e3}. 

\noindent Set
\begin{equation}
A_{o}(q^{-1})=\displaystyle{\prod_{k=0}^{\eta_{p}-1}}\left(1-p_{k}(q^{-1})\right)^{\frac{\eta_a}{\eta_p}}
\end{equation}

\noindent and consider $\hat{G}(q^{-1})=\frac{\sum_{k=1}^{\frac{\eta_a}{\eta_p}}\hat{n}_k^{T}V_{k}(q^{-1})}{1+\sum_{k=1}^{\frac{\eta_a}{\eta_p}}\hat{m}_k^{T}V_{k}(q^{-1})}$ \newline It is clear from \eqref{e3} that $V_{\frac{\eta_a}{\eta_p}}(q^{-1})$ has a characteristic polynomial equal to $A_{o}(q^{-1})$, and that we can perform a reduction to the same denominator of the two expressions  $\sum_{k=1}^{\frac{\eta_a}{\eta_p}}\hat{m}_k^{T}V_{k}(q^{-1})$ and $\sum_{k=1}^{\frac{\eta_a}{\eta_p}}\hat{n}_k^{T}V_{k}(q^{-1})$ (this denominator being $A_{o}$). Then one can write \[\hat{G}(q^{-1})=\frac{\sum_{k=1}^{\eta_a}\hat{b}_{k}q^{-k}}{1+\sum_{k=1}^{\eta_a}\hat{a}_{k}q^{-k}} \] which agrees with the classical output error model $y(t)=\frac{B(q^{-1})}{A(q^{-1})}u(t)+v(t)$, where $B(q^{-1})=\sum_{k=1}^{\eta_a}b_{k}q^{-k}$ and $A(q^{-1})=1+\sum_{k=1}^{\eta_a}a_{k}q^{-k}$. \newline

 \noindent Similarly, the stochastic part of the equation error model entails \\ \[ \hat{W}(q^{-1})=\frac{\sum_{k=1}^{\frac{\eta_a}{\eta_p}}\hat{l}_k^{T}V_{k}(q^{-1})}{1+\sum_{k=1}^{\frac{\eta_a}{\eta_p}}\hat{m}_k^{T}V_{k}(q^{-1})}=\frac{1+\sum_{k=1}^{\eta_a}\hat{c}_{k}q^{-k}}{1+\sum_{k=1}^{\eta_a}\hat{a}_{k}q^{-k}} \]Therefore the generalized ARMAX predictor agrees with the classical ARMAX model $y(t)=G(q^{-1})u(t)+W(q^{-1})e(t)$, with $G(q^{-1})=\frac{B(q^{-1})}{A(q^{-1})}$ and $W(q^{-1})=\frac{C(q^{-1})}{A(q^{-1})}$, where $C(q^{-1})=1+\sum_{k=1}^{\eta_a}\hat{c}_{k}q^{-k}$.  \newline

\noindent The generalized ARX predictor corresponds to $l_k=0$ in the expression of $\hat{W}$. Thus for the generalized ARX model we have $\hat{W}(q^{-1})=\frac{A_{o}(q^{-1})}{1+\sum_{k=1}^{\eta_a}a_{k}q^{-k}}$, and this predictor agrees with the model $y(t)=G(q^{-1})u(t)+W(q^{-1})e(t)$, \newline where  $W(q^{-1})=\frac{A_{o}(q^{-1})}{A(q^{-1})}$.

\section{Algorithms and their convergence conditions}
In the context of PLR, the predicted output is expressed from a linear combination of the estimated parameter vector $\hat\theta(t)$ and a regressor $\phi(t)$ following \eqref{e_yhat}.  \newline Let:  $\varepsilon(t+1)=y(t+1)-\hat{y}(t+1)$ be the prediction error. The estimation of model parameters is, most of the time, computed recursively, with the so-called parameter adaptation algorithm (PAA) \cite{r4}
\begin{subequations}
	\begin{align}
				\widehat{\theta } (t +1) =\widehat{\theta } (t) +F (t) \phi  (t) \varepsilon  (t +1)   \\
		    F^{ -1} (t +1) =\lambda _{1} F^{ -1} (t) +\lambda _{2} \phi  (t)\phi ^{T} (t)
		\end{align}
		\label{e5}
\end{subequations}
\noindent Where $F(t)$ is the adaptation gain (positive definite matrix),  and \\ $0<\lambda_{1}\leq1, 0\leq\lambda_{2}<2 $ the forgetting factors. \\

\noindent Each predictor is linked to an algorithm presented below. The generalized ARX is included in what we call the H-Recursive Least Square (H-RLS), (H stands for the Hambo transform which is associated with the bases used in this article). The generalized ARMAX predictor is associated with the H-Recursive Extended Least Squares (H-ERLS) algorithm, and the generalized Output Error predictor is the one used in what we call the H-Open-Loop Output Error algorithm (H-OLOE) in the sequel. 

\noindent It is well known that the algorithm convergence depends upon the strict real positiveness of a transfer transfer function appearing in the expression of the prediction error \cite{r4}, (chap. 3 and 4). For each algorithm we now present these convergence conditions, that partially differ from the convergence conditions of the classical algorithms established with the basis $q^{-1},q^{-2} \cdots$. Furthermore, we make use of recent results regarding Parameter Adaptation Algorithm (PAA) convergence \cite{r8}.

\subsection {Generalized ARX predictor, and H-RLS algorithm}
\label{sub_sec_ARMAX}
From \eqref{e_ARX}, \eqref{e_yhat}, and \eqref{e1}, one obtains immediately
\[\varepsilon(t+1)=\left(\theta_{0}-\theta\right)^{T}\phi(t)+e(t+1) \]
\noindent with:
\small{
\begin{multline*}
 {\phi^{T}(t)=\left[-V_1^{T}(q^{-1})y(t+1) \quad -V_2^{T}(q^{-1})y(t+1) \cdots \right.}  \\ 
{ \left. \cdots V_1^{T}(q^{-1})u(t+1) \quad V_2^{T}(q^{-1})u(t+1)  \cdots \right] }
\end{multline*}
}
\normalsize{
\noindent and:}
\small{
\[  \theta^{T}_{0}=[m_1^T \ m_2^T \cdots n_1^T \ n_2^T \cdots  ] \]
}
\normalsize
\noindent Exactly as for the classical recursive least-square there is no convergence condition.
 
\subsection {Generalized ARMAX predictor, and H-ERLS algorithm}
\noindent From \eqref{e_ARX}, \eqref{e_yhat}, and \eqref{e2}  in a deterministic context we have again
\[\varepsilon(t+1)=\left(\theta_{0}-\theta\right)^{T}\phi(t), \]

\small

\begin{multline*}
{ \phi^{T}(t)=\left[-V_1^{T}(q^{-1})y(t+1) \quad -V_2^{T}(q^{-1})y(t+1) \cdots  \right. } \\
{\cdots V_1^{T}(q^{-1})u(t+1) \quad  V_2^{T}(q^{-1})u(t+1)   \cdots   } \\
{\left. \cdots V_1^{T}(q^{-1})\varepsilon(t+1)  \quad  V_2^{T}(q^{-1})\varepsilon(t+1)  \cdots\right]  }
\end{multline*}
\small
\[ \theta^{T}_{0}=[m_1^T \ m_2^T \cdots n_1^T \ n_2^T \cdots  l_1^T \ l_2^T \cdots ] \]
\normalsize \noindent Therefore there is no convergence condition in a deterministic context. \newline

\noindent In a stochastic context, from \eqref{e_ARX}, \eqref{e_yhat}, and \eqref{e2}, we get easily
\begin{equation}
C(q^{-1})\varepsilon(t+1)=A_{o}\left(\theta_{0}-\theta\right)^{T}\phi(t) +C(q^{-1})e(t+1)
\label{eps_ARMAX}
\end{equation}

\noindent Notice that this expression generalizes the expression of the classical prediction error expression of extended-least squares algorithms corresponding to the case $A_{o}(q^{-1})=1$. \newline

\noindent  The convergence analysis of the H-ERLS algorithm in a stochastic context can be carried out by means of the martingale theory, and the recent results of \cite{r8}, generalizing those of theorem 4.2 in \cite{r4}. For this purpose, notice that $\{e(t)\}$ is a martingale difference sequence as defined in \cite{r4} p. 135, with
\begin{equation}
\mathbf{E}\left[e(t+1) | \mathcal{F}_{t}\right]=0 
\end{equation}
\begin{equation}
 \lim_{N\to\infty}sup\frac{1}{N}\sum_{t=1}^{N}e^{2}(t)<\infty 
\end{equation}

\noindent Where $\mathcal{F}_t$ is the $\sigma$-algebra of all observations generated up to $t$. \newline
\linebreak

\begin{theo}
\label{th2}
Consider the H-ERLS algorithm associated with the generalized ARMAX predictor in a stochastic context, and a prediction error as in \eqref{eps_ARMAX}. Assume that the following assumptions hold
\begin{enumerate}[a)]
\item The true system is in the model set,
\item \[lim_{N\to \infty} \frac{1}{N}\sum_{t=1}^{N}\phi^{T}(t)\phi(t)<\infty \]
\item $\lambda_1=1$
\item The transfer function
\[   \frac{A_0\left(z^{-1}\right)}{C\left(z^{-1}\right)} -\frac{\lambda_2}{2} \]
is strictly positive real (SPR). 
\end{enumerate}

\noindent Then one has
\begin{enumerate}
\item $\lim_{N\to\infty}\frac{1}{N}\sum_{t=1}^{N}\left[\varepsilon(t)-e(t)\right]^{2}=0  \quad a.s.$
\item $\lim_{N\to\infty}\frac{1}{N}\sum_{t=1}^{N}\varepsilon^{2}(t)=$\\$ \lim_{N\to\infty}\frac{1}{N^{1+\nu}}\sum_{t=1}^{N}e^{2}(t) \quad a.s.$
\item $\lim_{N\to\infty}\frac{1}{N}\sum_{t=1}^{N}\left[\left(\hat{\theta}-\theta_{0}\right)^T\phi(t-1)\right]^2=0 \quad a.s.$
\end{enumerate}
\noindent Moreover  if $\lim_{t\to\infty}F^{-1}(t)>0 \quad a.s. $ then 
\[\lim_{t\to\infty}\hat{\theta}(t)=\theta_{0} \quad a.s \]

\begin{proof}  The results are directly derived from theorem 4.2 of \cite{r4}. Condition b) ($\lim_{N\to\infty}\frac{1}{N}\sum_{t=1}^{N}\phi^{T}(t)\phi(t)<\infty$) is obtained from lemma 4.1 of \cite{r4}.  
\end{proof}

\end{theo}

\noindent Additionally the choice of the poles basis, $A_{o}(q^{-1})$ is depending on, can be used to relax the convergence condition of the H-ERLS algorithm in a stochastic context.

\subsection {Generalized Output Error predictor, and H-OLOE algorithm}

\noindent From \eqref{e_OE}, \eqref{e_yhat}, and \eqref{e2} we have
\begin{equation}
A(q^{-1})\varepsilon(t+1)=A_{o}(q^{-1})\left(\theta_{0}-\theta\right)^{T}\phi(t)+A(q^{-1})v(t+1)
\label{e_eps_OE}
\end{equation}

\noindent
where
\small
\begin{multline*}
  {\phi^{T}(t)=\left[-V_1^{T}(q^{-1})\hat{y}(t+1) \quad -V_2^{T}(q^{-1})\hat{y}(t+1) \cdots  \right. } \\
	{\cdots \left. V_1^{T}(q^{-1})u(t+1) \quad V_2^{T}(q^{-1})u(t+1)  \cdots \right] }
\end{multline*}
\normalsize

\small
\[  \theta^{T}_{0}=[m_1^T \ m_2^T \cdots n_1^T \ n_2^T \cdots  ] \]
\normalsize

\noindent Hence the following theorem is obtained from \cite{r8}: \newline

\noindent In a stochastic context, if $v(t)=e(t)$ (meaning that the output noise is a white noise and therefore a martingale difference sequence), one has the following result:  \newline

\begin{theo}
\label{th3}
Consider the H-OLOE algorithm and the associated generalized output error predictor in a stochastic context, and its prediction error given by \eqref{e_eps_OE} where $\{v(t)\}$ is a white noise ($v(t)=e(t)$). Assume that the following assumptions hold\\
\begin{enumerate}[a)]
\item Assume that the stationary processes $\phi(t,\hat{\theta})$ and $\varepsilon(t+1,\hat{\theta})$ can be defined for $\hat{\theta}(t)=\theta_0$
\item Assume that $\hat{\theta}(t)$ generated by the algorithm belongs infinitely often to the domain $D_s$ for which the stationary processes $\phi(t,\hat{\theta})$ and $\varepsilon(t+1,\hat{\theta})$ can be defined
\item Define the convergence domain $D_c$ such that
\[D_c:[\theta: \phi^T(t,\theta)[\theta_0-\theta]]=0 \]
\item $\lambda_1=1$
\item If the transfer function
\[   \frac{A_0\left( z^{-1}\right)}{A\left( z^{-1}\right)} -\frac{\lambda_2}{2} \]
is SPR. 
\end{enumerate}

\noindent Then
\[Prob\{lim_{t \to \infty} \hat{\theta}(t)\in D_c \}=1 \]

\begin{proof}  Directly issued from theorem 4.1 of \cite{r4}.
\end{proof}

\end{theo} 

\noindent If $v(t)$ is not a white noise, the convergence of the algorithm can be proved  for $\lambda=1$, provided the transfer function $\frac{A_o(z^{-1})}{A(z^{-1})} -\frac{\lambda_2}{2}$ is SPR with theorem 4.1 of \cite{r4}.

\noindent Remark that the choice of the poles basis, $A_{o}(q^{-1})$ is depending on,  can be used to relax the convergence condition of the H-OLOE algorithm, both in a deterministic or stochastic context.

\section{Bias distribution analysis}

\subsection{Limit models expressions}
The output of the true system can be expressed as
\[y(t)=G(q)u(t)+W(q)e(t) \]

\noindent On the other hand, the stationary condition of the parameter adaptation algorithm is
\begin{equation}
\mathbf{E}\left[\varepsilon(t+1)\phi(t,{\theta})\right]=0
\label{e13}
\end{equation}
\noindent  This limit exists for a strictly decreasing adaptation gain $F(t)$, i.e. for $\lambda_{1}=1$. One assumes in this section that $\lambda_{1}=1$. Except the case of least squares algorithm, the regressor $\phi(t,{\theta})$ depends on the estimated parameters. As shown in \cite{r9}, condition  \eqref{e13} does not imply in general the minimization of $\mathbf{E}\left[\varepsilon^{2}(t)\right]$ (particularly if the system is not in the model set). This is the major difference with prediction error methods (PEM) that aim directly at minimizing this  latter expression. Thus it is important to determine the signal whose variance is effectively minimized if the condition \eqref{e13} is satisfied, in order to infer the effective bias distribution in the frequency domain.  As in \cite{r9}, let us denote by $\varepsilon_{E}(t+1,{\theta})$ the equivalent prediction error signal (in general non measurable) such that the optimal estimated parameters vector $\hat{\theta}^{*}$ of PLR algorithms is given by 
\begin{equation}
\hat{\theta}^{*}=Argmin\mathbf{E}\left[\varepsilon_{E}^{2}(t+1,\theta)\right]
\label{e11} 
\end{equation}

\noindent It is shown in \cite{r9}, that for the equation error model one has
\begin{equation}
\varepsilon_{E}(t+1,\theta)=Q(q^{-1},{\theta})\varepsilon(t+1,\theta)+(1-Q(q^{-1},{\theta}))e(t+1)
\end{equation}
\noindent and for the output error model
\begin{equation}
\varepsilon_{E}(t+1,\theta)=Q(q^{-1},{\theta})\varepsilon(t+1,\theta)+(1-Q(q^{-1},{\theta}))v(t+1)
\end{equation}
Where
\noindent $Q(q^{-1},{\theta})\frac{\partial\varepsilon(t+1,{\theta})}{\partial{\theta}}=-\phi(t,{\theta})$. \newline

\noindent Consequently we infer the two following theorems:

\begin{theo}
\label{th2b}
The equivalent prediction error signal for the H-ERLS algorithm associated with the generalized-ARMAX predictor is given by
\begin{equation}
\varepsilon_{E}(t)=\frac{\widehat{A}}{A_{0}}\left[\left(G-\hat G\right)u(t)+\left(W-\frac{\widehat C}{\widehat A} \right)e(t) \right]+e(t)
\label{e14}
\end{equation}
\begin{proof}
For the ARMAX predictor, one has $Q(q^{-1},\theta)\frac{\partial\varepsilon(t+1)}{\partial\theta}=-\phi(t)$ with  \newline $Q(q^{-1},\theta)=1+\sum_{k=1}^{\frac{\eta_a}{\eta_p}}\hat l_{k}^{T}V_{k}(q^{-1})=\frac{\widehat{C}}{A_{o}}$, and owing to theorem 1 of \cite{r9}, $\varepsilon_{E}(t+1)=Q(q^{-1},\theta)\varepsilon(t+1)+(1-Q(q^{-1},\theta))e(t+1)$, that yields the result.

\end{proof}
\end{theo}

\begin{theo}

The equivalent prediction error of the H-OLOE algorithm, associated with the generalized-output error predictor is given by 
\begin{equation}
\varepsilon_{E}(t)=\frac{\widehat{A}}{A_{0}}\left[\left(G-\hat G\right)u(t) \right]+v(t)
\label{e15}
\end{equation}
\begin{proof}
For the output error predictor, we have $Q(q^{-1},\theta)\frac{\partial\varepsilon(t+1)}{\partial\theta}=-\phi(t)$, \newline with $Q(q^{-1},\theta)=1+\sum_{k=1}^{\frac{\eta_a}{\eta_p}}\hat m_{k}^{T}V_{k}(q^{-1})=\frac{\widehat{A}}{A_{o}}$; once again by applying theorem 1 of \cite{r9}, we have that $\varepsilon_{E}(t+1)=Q(q^{-1},\theta)\varepsilon(t+1)+(1-Q(q^{-1},\theta))v(t+1)$, leading to expression \eqref{e15}.
\end{proof}
\end{theo}

\noindent Additionally, one checks immediately that for the H-RLS algorithm corresponding to the generalized-ARX predictor, since the regressor $\phi(t)$ is independent of $\hat\theta(t)$, the prediction error and the equivalent prediction error are equal and 
\begin{equation}
\varepsilon_{E}(t)=\varepsilon(t)=\frac{\widehat{A}}{A_{o}}\left[\left(G-\widehat{G}\right)u(t)+\left(W-\frac{A_o}{\widehat{A}}\right)\right]+e(t)
\label{e16}
\end{equation}

\noindent From \eqref{e14}, \eqref{e15}, \eqref{e16}, one can infer the limit models expressed in Table. \ref{tb2},

\begin{table}[H]
\begin{center}
\begin{tabular}{|>{\centering\tiny\arraybackslash}p{35pt}|>{\centering\small\arraybackslash}p{193pt}|}
\hline

ALGORITHMS &  {${\widehat{\theta}}^*$}  \\
\hline

 \scriptsize{H-RLS (generalized ARX predictor)} & \scriptsize{ $  Argmin \displaystyle{\int\nolimits_{-\pi}^{+\pi}}  \left\{\left|\frac{ \widehat{A} (e^{i\omega})}{\widehat{A}_{o} (e^{i\omega})} \right |^2\left(   \left|G(e^{i\omega})- \widehat{G} (e^{i\omega}) \right|^2 {\Phi}_{uu} (\omega) \right. \right. \newline \left.\left. \cdots +\left|W(e^{i\omega})-\frac{\widehat{A}_{o}(e^{i\omega})}{\widehat{A}(e^{i\omega})} \right|^2 {\Phi}_{ee}(\omega) \right)\right\} \mathrm d\omega  $ }\\ 
\hline
 \scriptsize{H-ERLS (Generalized-ARMAX predictor)} &\scriptsize{$ Argmin \displaystyle\int\nolimits_{-\pi}^{+\pi}  \left\{\left| \frac{ \widehat{A} (e^{i\omega})}{\widehat{A}_{o} (e^{i\omega})}  \right |^2 \left( \left|G(e^{i\omega})- \widehat{G} (e^{i\omega}) \right|^2 {\Phi}_{uu} (\omega)\right.\right. \newline \cdots  \left.\left. +\left|W(e^{i\omega})-\frac{\widehat{C}(e^{i\omega})}{\widehat{A}(e^{i\omega})} \right|^2 {\Phi}_{ee}(\omega)\right)\right\} \mathrm d\omega $ }\\ 
\hline
  \scriptsize{H-OLOE (Generalized-OUTPUT ERROR predictor)} & {\quad} \newline 
	{\scriptsize{ $ Argmin \displaystyle\int\nolimits_{-\pi}^{+\pi}  \left| \frac{ \widehat{A} (e^{i\omega})}{\widehat{A}_{o} (e^{i\omega})} \right |^2  \left|G(e^{i\omega})- \widehat{G} (e^{i\omega}) \right|^2 {\Phi}_{uu} (\omega)  \mathrm d\omega $}} \\ 
\hline

 \end{tabular}
\caption{Limit model expressions for open-loop PLR algorithms including predictors expressed with generalized orthonormal transfer functions}
\label{tb2}
\end{center}
\end{table}

\noindent where $\Phi_{uu}(\omega), \Phi_{ee}(\omega)$, are the spectral density associated with respectively $\{u(t)\}$ and $\{e(t)\}$. \newline

\noindent The results in table \ref{tb2}, lead to some remarks:
\begin{itemize}
\item The bias distribution of algorithms parameterized with generalized orthonormal functions differ from standard PLR algorithms.
\item The bias distribution depends on the basis poles. Therefore these poles can be considered as tuning parameters in order to adjust the model fit over the frequency domain. This dependance is a direct consequence of the results of \cite{r9}, and are analyzed with the concept of equivalent prediction error. Note that this dependence is specific to PLR algorithms and would not apply to PEM schemes, for which a parameterization modification has no effect on the identified model (see \cite{r11} p. 437).
\item The limit expressions in Table \ref{tb2} depend all on the same weighting function: $\left|\frac{\widehat{A}(e^{i\omega})}{A_o(e^{i\omega})} \right|^2$, consequently there is a homogeneity in the effect due to the basis poles, independently of the predictor structure.
\item For Output-Error and ARMAX predictors based schemes, the noise model is not affected by the the parameterization (contrary to a classical prediction error filtering applied on standard schemes, that modify the noise model, cf. \cite{r10}, \cite{r11}).
\end{itemize}

\subsection{Effect of the basis poles on the bias distribution}

\noindent From \eqref{e16}, one can write for  H-RLS
\begin{equation}
\varepsilon_{E}(t)=\frac{\hat{A}(q)}{A_0(q)}y(t)-\frac{\hat{B}(q)}{A_0(q)}u(t)
\label{eq_RLS}
\end{equation}

\noindent Set $y_d(t)=G_0(q)u(t)$. For H-OLOE, from \eqref{e15}  one has
\begin{equation}
\varepsilon_{E}(t)=\frac{\hat{A}(q)}{A_0(q)}y_d(t)-\frac{\hat{B}(q)}{A_0(q)}u(t)+v(t)
\label{eq_OLOE}
\end{equation}
\noindent  And for H-ERLS, from \eqref{e14}
\begin{equation}
\varepsilon_{E}(t)=\frac{\hat{A}(q)}{A_0(q)}y(t)-\frac{\hat{B}(q)}{A_0(q)}u(t)-\frac{\hat{C}(q)}{A_0(q)}e(t)+e(t)
\label{eq_ERLS}
\end{equation}

\noindent Consider the expression $\frac{\hat{A}(q)}{A_0(q)}=1+\sum_{k=1}^{\frac{\eta_a}{\eta_0}}{\hat{m}}_k^{T}V_{k}(q)$. From equations 12.20, 12.30 and 12.31 of  \cite{r1}, its Hambo (operator) transform $\tilde{\mathcal{\hat{A}}}(\lambda)$ can be written
\begin{equation}
\tilde{\mathcal{\hat{A}}}(\lambda)=\mathscr{\hat{A}}_{0}+\mathscr{\hat{A}}_{1}\lambda^{-1}+ \cdots +\mathscr{\hat{A}}_{\eta_a/\eta_p}\lambda^{-{\eta_a/\eta_p}}
\label{eq_M}
\end{equation}

\noindent where $\mathscr{\hat{A}}_i \in \mathbb{R}^{\eta_p \times \eta_p}$. \newline
\noindent Similarly the Hambo (operator) transform of  $\sum_{k=1}^{\frac{\eta_a}{\eta_p}}{\hat{n}}_k^{T}V_{k}(q) =\frac{\hat{B}(q)}{A_0(q)}$, denoted $\tilde{\mathcal{\hat{B}}}(\lambda)$ can be put under the form
\begin{equation}
\tilde{\mathcal{\hat{B}}}(\lambda)=\mathscr{\hat{B}}_0+\mathscr{\hat{B}}_1\lambda^{-1}+ \cdots +\mathscr{\hat{B}}_{\eta_a/\eta_p}\lambda^{-{\eta_a/\eta_p}}
\label{eq_N}
\end{equation}
\noindent where $\mathscr{\hat{B}}_i \in \mathbb{R}^{\eta_p \times \eta_p}$. \newline
\noindent and the Hambo operator transform of $1+\sum_{k=1}^{\frac{\eta_a}{\eta_p}}{\hat{l}}_k^{T}V_{k}(q) =\frac{\hat{C}(q)}{A_0(q)}$, denoted $\tilde{\mathcal{\hat{C}}}(\lambda)$ can be expressed as
\begin{equation}
\tilde{\mathcal{\hat{C}}}(\lambda)=\mathscr{\hat{C}}_0+\mathscr{\hat{C}}_1\lambda^{-1}+ \cdots +\mathscr{\hat{C}}_{\eta_a/\eta_p}\lambda^{-{\eta_a/\eta_p}}
\end{equation}
\noindent where $\mathscr{\hat{C}}_i \in \mathbb{R}^{\eta_p \times \eta_p}$. \newline

\noindent Note that $\tilde{\mathcal{\hat{A}}}(\lambda)$, $\tilde{\mathcal{\hat{B}}}(\lambda)$, $\tilde{\mathcal{\hat{C}}}(\lambda)$ are "FIR" filters in the Hambo domain. \newline

\noindent Let us consider the Hambo transform of $\{\varepsilon_{E}(t)\}$, denoted $\{\tilde{\varepsilon}(\lambda)\}$, and $\{\tilde{y}(\lambda)\}$,$\{\tilde{u}(\lambda)\}$, $\{\tilde{e}(\lambda)\}$, $\{\tilde{y_d}(\lambda)\}$,$\{\tilde{v}(\lambda)\}$
the Hambo transforms of $\{y(t)\}$, $\{u(t)\}$, $\{e(t)\}$, $\{y_d(t)\}$,$\{v(t)\}$ respectively (defined from \cite{r1} chap. 12). From \eqref{eq_RLS}, one obtains

\begin{equation}
\tilde{\varepsilon}(\lambda)=\tilde{\mathcal{\hat{A}}}(\lambda)\tilde{y}(\lambda)-\tilde{\mathcal{\hat{B}}}(\lambda)\tilde{u}(\lambda)
\label{eq_err_RLS}
\end{equation}

\noindent But, since $\tilde{\mathcal{\hat{A}}}(\lambda)$, $\tilde{\mathcal{\hat{B}}}(\lambda)$ are "FIR" filters, equation \eqref{eq_err_RLS} is nothing else than an equation error in the Hambo domain (note that $\tilde{y}(\lambda)$ and $\tilde{u}(\lambda)$  do not depend on the estimated parameters). \newline

\noindent For the H-OLOE scheme, from \eqref{eq_OLOE} one obtains what we call a "pseudo error equation" ( $\tilde{y}_d(\lambda)$ that appears here is non-measurable but independent with respect to the estimated parameters)
\begin{equation}
\tilde{\varepsilon}(\lambda)=\tilde{\mathcal{\hat{A}}}(\lambda)\tilde{y_d}(\lambda)-\tilde{\mathcal{\hat{B}}}(\lambda)\tilde{u}(\lambda)+\tilde{v}(\lambda)
\label{eq_err_OLOE}
\end{equation}

\noindent and for the H-ERLS scheme, from \eqref{eq_ERLS} one has another pseudo error equation ($\tilde{e}(\lambda)$ is non-measurable but independent with respect to the estimated parameters)
\begin{equation}
\tilde{\varepsilon}(\lambda)=\tilde{\mathcal{\hat{A}}}(\lambda)\tilde{y}(\lambda)-\tilde{\mathcal{\hat{B}}}(\lambda)\tilde{u}(\lambda)-\tilde{\mathcal{\hat{C}}}(\lambda)\tilde{e}(\lambda)+\tilde{e}(\lambda)
\label{eq_err_ERLS}
\end{equation}

\bigskip

\noindent Let us call $\tilde{\mathcal{G}}(\lambda)$, $\tilde{\mathcal{\hat{G}}}(\lambda)$, and $\tilde{\mathcal{W}}(\lambda)$ the Hambo signal transforms of $G(q)$, $\hat{G}(q)$ and $W(q)$ respectively.  \newline

\noindent For H-RLS, from \eqref{eq_err_RLS}  one has
\begin{equation}
\tilde{\varepsilon}_E(\lambda)=\tilde{\mathcal{\hat{A}}}(\lambda)\left( \tilde{G}(\lambda)- \tilde{\hat{G}}(\lambda)\right)\tilde{u}(\lambda)+\tilde{\mathcal{\hat{A}}}(\lambda)\left(\tilde{W}(\lambda)-\tilde{\mathcal{\hat{A}}}^{-1}(\lambda)\right)\tilde{e}(\lambda)+\tilde{e}(\lambda)
\label{eq_err_RLS2}
\end{equation}

\noindent For H-OLOE, from \eqref{eq_err_OLOE}   one can write
\begin{equation}
\tilde{\varepsilon}_E(\lambda)=\tilde{\mathcal{\hat{A}}}(\lambda)\left( \tilde{G}(\lambda)- \tilde{\hat{G}}(\lambda)\right)\tilde{u}(\lambda)+\tilde{v}(\lambda)
\label{eq_err_OLOE2}
\end{equation}

\noindent And for H-ERLS, from \eqref{eq_err_ERLS}  one has
\begin{equation}
\tilde{\varepsilon}_E(\lambda)=\tilde{\mathcal{\hat{A}}}(\lambda)\left( \tilde{G}(\lambda)- \tilde{\hat{G}}(\lambda)\right)\tilde{u}(\lambda)+\tilde{\mathcal{\hat{A}}}(\lambda)\left(\tilde{W}(\lambda)-\tilde{\mathcal{\hat{A}}}^{-1}(\lambda)\tilde{\mathcal{\hat{C}}}(\lambda)\right)\tilde{e}(\lambda)+\tilde{e}(\lambda)
\label{eq_err_ERLS2}
\end{equation}

\noindent Remark that these expressions are exactly the same as those of native PLR schemes given in Table 2 of \cite{r9}, but expressed now with the Hambo operator! \newline

\noindent On the other hand, the Parseval equality holds in the Hambo domain (see \cite{r2}), and one can write
\begin{equation}
 \mathbf{E}[\varepsilon_{E}^{2}(t)]=\frac{1}{2\pi i} \oint_{\mathbf{T}} \tilde{\varepsilon}_E^{T}(\lambda)\tilde{\varepsilon}_E(\lambda^{-1})\frac{d\lambda}{\lambda} 
\label{Parseval_L}
\end{equation}

\noindent The Hambo frequency $\omega_{\lambda}$ is defined  such that $\lambda=e^{i\omega_{\lambda}}$, where $\omega_{\lambda} \in \left[-\pi\eta_{p}, +\pi\eta_{p}\right]$.
The relation between $\omega$ and $\omega_{\lambda}$ has been first studied in \cite{r12}, with the introduction of the phase function called the $\beta$ function (see \cite{r1}, p.222), which is bijective and strictly increasing, and where $\omega_{\lambda}=\beta(\omega)$. In particular one has 
\begin{equation}
d\omega_{\lambda}=\beta{'}(\omega)d\omega
\end{equation} 
\noindent  and as shown in the same reference
\begin{equation}
 \beta^{'}(\omega)=V_{1}^{T}(e^{i\omega})V_{1}(e^{-i\omega})
\label{e98b}
\end{equation}

\noindent From \eqref{Parseval_L} one can write \footnote{The interest of $\omega_\lambda$ expressed over $\left[-\pi\eta_{p}, +\pi\eta_{p}\right]$ is to circumvent the issue due to the fact that the mapping from $\lambda$ to z is one to $\eta_p$.}
\begin{equation}
\mathbf{E}[\varepsilon_E^{2}(t)]=\frac{1}{2\eta_p\pi}\int_{-\eta_p\pi}^{\eta_p\pi}\tilde{\varepsilon}_{E}^{T}(e^{i\omega_{\lambda}})\tilde{\varepsilon}_{E}(e^{-i\omega_{\lambda}})d\omega_{\lambda}
\label{eq_pars}
\end{equation}





\noindent From \eqref{eq_err_RLS2}, \eqref{eq_err_OLOE2}, \eqref{eq_err_ERLS2}, one can express limit models for PLR schemes based on GBOF exactly similar to those of native PLR schemes given in Table 3 of \cite{r9}, but now with Hambo operators and spectral densities in the Hambo domain on the distorted $\omega_{\lambda}$ frequency scale !  \newline


\noindent One can infer that the analysis of this distortion provides insights about bias distribution: The frequencies $\omega$  for which the distortion (or dilatation) rate from $\omega$ scale to $\omega_{\lambda}$ scale is maximum, are over-penalized in the criterion minimization (we can expect better model fit around these frequencies), whereas the frequencies corresponding to a low dilatation are under-weighted (inducing a worse model approximation).   \newline

\subsection{A heuristic method for evaluating the effect of the basis poles on the bias distribution}

If we consider the relation from $\omega$ to $\omega_{\lambda}$ by means of the $\beta(\omega)$ function, the frequencies $\omega$  for which the distortion (or dilatation) rate from $\omega$ scale to $\omega_{\lambda}$ scale is maximum, are over-penalized in the criterion minimization (we can expect better model fit around these frequencies), whereas the frequencies corresponding to a low dilatation are under-weighted (inducing a worse model approximation). Then the frequency distortion analysis  from $\omega$ to $\omega_\lambda$ scales, gives an useful indication about the effect of the basis poles on the model fit quality.  \newline

\noindent However as most of linear systems are represented in Bode diagrams with a logarithmic scale such that $\bar\omega=log(\omega)$, it is more interesting to study the dilatation (or distortion) rate from $\bar{\omega}$ to $\omega_{\lambda}$.			
The relation between measures of integration is
\begin{equation}
d\omega_{\lambda}=e^{\bar{\omega}}\beta^{'}(e^{\bar\omega})d\bar{\omega}
\label{e20}
\end{equation}
	
 \noindent According to \cite{r12}, and \cite{r1} p. 222, one has
\begin{equation}  
\beta^{'}(\omega)=\sum_{k=0}^{\eta_{p}-1}\beta^{'}_{k}(\omega) 
\end{equation}  

\begin{equation}
 \beta^{'}_{k}(\omega)=\frac{1-\lvert p_{k}\rvert^2}{\lvert1-\bar{p}_{k}e^{i\omega}\rvert^2} 
\end{equation}

\noindent  Remark that $\beta^{'}(\omega)$ is nothing but a particular expression of the reproducing Kernel of the associated orthogonal transfer function basis, see \cite{r14} (chap.4).
\noindent Equation \eqref{e20} leads to define the distortion rate function $\chi(\omega)$ from $\bar\omega$ scale to $\omega_{\lambda}$ scale, such that
\begin{equation}
\chi(\omega)=\frac{1}{\pi}\omega\beta^{'}(\omega)={\frac{1}{\pi}}e^{\bar\omega}\beta^{'}(e^{\bar\omega})
\end{equation}

\noindent The following property \ref{prop6}, corresponds to a conservation principle of $\chi(\omega)$

\begin{property}
\label{prop6}
One has
\begin{equation}
\int_{-\infty}^{log(\pi)}\chi(e^{\bar{\omega}})d\bar{\omega}=1 
\end{equation}
\begin{proof}
One has $\beta^{'}(\omega)=V_{1}^{T}(e^{i\omega})V_{1}(e^{i\omega})$, and because of the orthonormality of $V_{1}(e^{i\omega})$ we have the result immediately, see \cite{r1},  p.88.
\end{proof}
\end{property}

\noindent Consider $\chi_k(\omega)=\frac{1}{\pi}\omega\beta^{'}_k(\omega)$ and define the k-th basis pole from its proper frequency $\omega_{ok}$ and its damping $\zeta_{k}$, such that $p_k=\rho_{k}e^{i\sigma_k}$ with $\rho_k=e^{-\zeta\omega_{ok}}$ and $\sigma_k=\sqrt{1-\zeta_k^{2}}\omega_{ok}$. Let $0<\rho_{k}<1$. This function $\chi_k(\omega)$ has nice properties presented in the following theorems: \newline

\begin{theo}
\label{theo4}
Assume $\zeta_k^{2}\geq 1-\frac{\pi^2}{4\omega_{ok}^{2}}$. One has the following results
\begin{enumerate}
\item If $cosh(\zeta_k\omega_{ok})-\sqrt{1-\zeta^{2}_{k}}\omega_{ok}\geq\frac{\pi}{2}$, $\chi_k(\omega)$ is an increasing function on $\left[0, \pi \right]$, and has its maximum at $\omega=\pi$.
\item If $cosh(\zeta_k\omega_{ok})-\sqrt{1-\zeta^{2}_{k}}\omega_{ok}<\frac{\pi}{2}$, $\chi_k(\omega)$ has a unique maximum on $\left[0, \pi \right[$. Additionnaly if: \newline
$cosh(\zeta\omega_{ok})+cos(\sqrt{1-\zeta^{2}_{k}}\omega_{ok})-\pi sin(\sqrt{1-\zeta^{2}_{k}}\omega_{ok}) >0$, $\chi(\omega)$ has necessarily a minimum. 
\item If $p_k$ is real ($\zeta_k=1$), and if $\frac{\pi-\sqrt{\pi^2-4}}{2}<p_k<1$, $\chi_k(\omega)$ has a unique maximum on $\left[0, \pi \right[$, and a unique minimum. If $p_k\leq\frac{\pi-\sqrt{\pi^2-4}}{2}$, $\chi_k(\omega)$ is an increasing function on $\left[0, \pi \right]$, and has its maximum at $\omega=\pi$.
\end{enumerate}

\begin{proof}
One has 
\[ \chi _{k}(\omega ) =\frac{1}{\pi }\frac{1 -\left \vert p_{k}\right \vert ^{2}}{\left \vert 1 -\overline{p}_{k}e^{i\omega }\right \vert ^{2}}\omega =\frac{1}{\pi }\frac{(1 -\rho _{k}^{2})}{1 +\rho _{k}^{2} -2\rho _{k}cos(\omega  -\sigma _{k})}\omega \] and
\[ \frac{ \partial \chi _{k}(\omega )}{ \partial \omega } =\frac{1}{\pi }\left (1 -\rho _{k}^{2}\right )\frac{\frac{1 +\rho _{k}^{2}}{2\rho_{k}} -\cos (\omega  -\sigma _{k}) -\omega \sin (\omega  -\sigma _{k})}{\left(\frac{1 +\rho _{k}^{2}}{2\rho_k} -\cos (\omega  -\sigma _{k})\right)^2} \]. \newline
\noindent   The sign of  $\frac{ \partial \chi _{k}(\omega )}{ \partial \omega }$ depends upon the sign of \newline $g(\omega)=\frac{1 +\rho _{k}^{2}}{2 \rho_{k}} - \cos (\omega  -\sigma _{k}) - \omega \sin (\omega  -\sigma _{k}) $.  
One has  \newline $\frac{\partial g(\omega)}{\partial \omega}=-\omega cos(\omega-\sigma_k)=-\omega cos(\omega -\sqrt{1-\zeta_k^{2}}\omega_{ok})$, $\frac{\partial g(\omega)}{\partial \omega} \leq0$ if and only if $\omega \leq \frac{\pi}{2}+\sqrt{1-\zeta^{2}_{k}}\omega_{ok}$, and $\frac{\partial g(\omega)}{\partial \omega} >0$ otherwise. \newline
\noindent Set $\breve{\omega}$ the frequency for which $g(\omega)$ is minimum. One has \newline
$g(\breve{\omega})=cos(\zeta\omega_{ok})-\sqrt{1-\zeta^{2}_{k}}\omega_{ok}-\frac{\pi}{2}$. \newline
\noindent This quantity is strictly negative if and only if
\begin{equation}
  cosh(\zeta_k\omega_{ok})-\sqrt{1-\zeta^{2}_{k}}\omega_{ok}<\frac{\pi}{2}
\label{ee500}
\end{equation}
\noindent Additionally $g(0)=cosh(\zeta_k\omega_{ok})-cos(\sqrt{1-\zeta_k^2}\omega_{ok})>0$ for any $\omega_{ok}>0$, $g(\omega)$ is a positive decreasing function for $\omega$ close to $0$, and has a minimum at $\omega=\frac{\pi}{2}+\sqrt{1-\zeta^{2}_{k}}\omega_{ok}$ only if $\omega_{ok}<\frac{\pi}{2\sqrt{1-\zeta_k^{2}}}$, i.e
\begin{equation}
 \zeta_k^{2}\geq 1-\frac{\pi^2}{4\omega_{ok}^{2}} 
\label{ee501}
\end{equation} 
  Since
\noindent  $g(\pi)=cosh(\zeta_k\omega_{ok})+cos(\sigma_k)-\pi sin(\sigma_k)$, \newline
\noindent Therefore $g(\pi)>0$ is equivalent to\newline
\begin{equation}
 cosh(\zeta\omega_{ok})+cos(\sqrt{1-\zeta^{2}_{k}}\omega_{ok})-\pi sin(\sqrt{1-\zeta^{2}_{k}}\omega_{ok}) >0 
\label{ee502}
\end{equation}
\noindent Therefore, if we assume that \eqref{ee501}  and \eqref{ee500} are satisfied, $\chi(\omega)$ has a unique maximum on $\left[0, \pi \right[$.  Furthermore if \eqref{ee502} is satisfied  $\chi(\omega)$ has  a unique minimum. \newline
\noindent If condition \eqref{ee501} is satisfied and \eqref{ee500} is not, $\chi(\omega)$ is an increasing function on this interval and admits a unique maximum at $\omega=\pi$.

\noindent If the pole $p_k$ is real condition \eqref{ee502} is necessarily fulfilled, and  \eqref{ee500} reduces to: \newline
 $ 1+p_k^{2}-\pi p_k<0 $\newline
\noindent Since we consider only stable poles, this is equivalent to \newline   $ p_k>\frac{\pi-\sqrt{\pi^2 -4}}{2} $.
\end{proof}

\end{theo}

\begin{figure}[H]
\begin{center}
\includegraphics[ width=3.4in, keepaspectratio ]{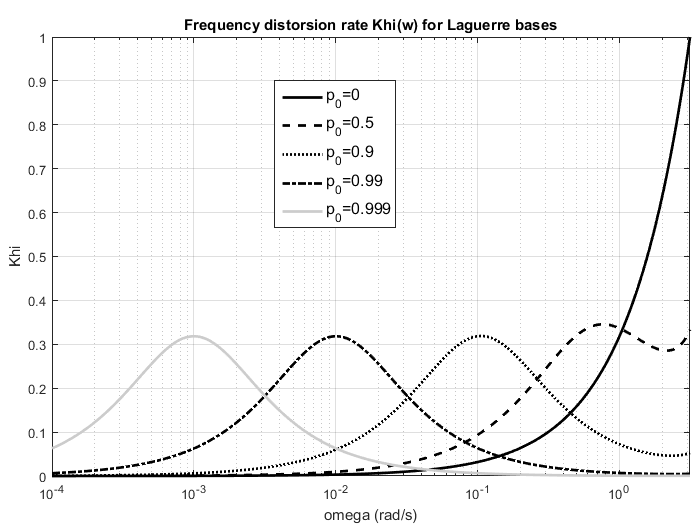}
\caption{Frequency distortion rate $\chi(\omega)$ for Laguerre bases}
\label{fig1}
\end{center}
\end{figure}

\begin{theo}
\label{theo5}
Set $\omega_{max}$ the frequency for which $\chi_{k}(\omega)$ is maximum. If $\omega_{ok}\rightarrow 0$, one has 
\begin{equation}
 \omega_{max}=\omega_{ok}+o\left(\left\vert\omega_{ok} \right\vert\right)
\end{equation}

\begin{proof}
According to theorem \ref{theo4}, $\omega_{max}$ is the smallest frequency such that $g(\omega)=0$. This frequency is such that \newline

$h(\omega_{ok},\omega)= cosh(\zeta\omega_{ok})-cos\left(\omega-\sqrt{1-\zeta^{2}_{k}}\omega_{ok}\right) \cdots$ \newline
$\cdots -\omega cos\left(\omega-\sqrt{1-\zeta^{2}_{k}}\omega_{ok}\right)$. \newline
Let us consider $\omega_{ok}$ as the function variable and $\omega$ as a parameter. One has \newline
\scriptsize{
$h(\omega_{ok},\omega)= - \left(cos(\omega)cos\left(\sqrt{1-\zeta^{2}_{k}}\omega_{ok}\right)+sin(\omega)sin\left(\sqrt{1-\zeta^{2}_{k}}\omega_{ok}\right)\right) \cdots$ \newline
$ -\omega\left(sin(\omega)cos\left(\sqrt{1-\zeta^{2}_{k}}\omega_{ok}\right)-cos(\omega)sin\left(\sqrt{1-\zeta^{2}_{k}}\omega_{ok}\right)\right) \cdots$. \newline
$ +cosh(\zeta\omega_{ok})$}.  \newline
\normalsize
A first order Taylor-Young approximation yields \newline
$h(\omega_{ok},\omega)=1-cos(\omega)+sin(\omega)\sqrt{1-\zeta^{2}_{k}}\omega_{ok}-\omega sin(\omega)+\omega cos(\omega)\sqrt{1-\zeta^{2}_{k}}\omega_{ok}+o(\omega_{ok})$. \newline
This quantity can be null only if $1-cos(\omega)-\omega sin(\omega)=0$, implying $\omega=0$. \newline
If we perform a second order Taylor-Young expansion near 0, we get \newline
\noindent $h(\omega_{ok},\omega)= 1+\frac{\zeta_k^{2}\omega_{ok}^2}{2}-\left(1-\frac{\omega^2}{2}\right)\left(1-\frac{(1-\zeta_{k}^2)\omega_{ok}^{2}}{2}\right)-\omega \sqrt{1-\zeta^{2}_{k}}\omega_{ok}  -\omega\left(\omega-\sqrt{1-\zeta^{2}_{k}}\omega_{ok}\right) +o\left(\left\Vert(\omega_{ok},\omega \right)\Vert\right)$ \newline
$=1+\frac{\zeta_{k}^{2}\omega_{ok}^2}{2}-1+\frac{(1-\zeta_{k}^2)\omega_{o}^2}{2}+\frac{\omega^2}{2}-\omega \sqrt{1-\zeta^{2}_{k}}\omega_{ok}-\omega^2+ \omega \sqrt{1-\zeta^{2}_{k}}\omega_{ok} +o^{2}\left(\left\Vert(\omega_{ok},\Omega(\omega_{ok}))\right\Vert\right) $ \newline

\noindent $=\frac{1}{2}\left(\omega^{2}-\omega_{ok}^{2}\right)+o^{2}\left(\left\Vert(\omega_{ok},\Omega(\omega_{ok})\right)\Vert\right)$ \newline

\noindent Consequently $h(\omega_{ok},\omega)=o^{2}(\left\Vert(\omega_{ok},\omega \right)\Vert)$ if and only if $\omega^{2}=\omega_{ok}^{2}$. \newline
\noindent The relation $h(\omega_{ok},\omega)=0$ entails an implicit function $\omega_{max}=\Omega(\omega_{ok})$, and one has:  \newline 
$h\left(\omega_{ok},\Omega(\omega_{ok})\right)=o^{2}\left(\left\Vert\left(\omega_{ok},\Omega(\omega_{ok})\right)\right\Vert\right)$. Hence the result.

\end{proof}

\end{theo}
\noindent For a one pole basis (Laguerre basis), if $p_{0}$ is sufficiently close to 1, the maximum of $\chi(\omega)$ corresponds to a frequency $\omega_{max}\approx \omega_{o0}$, and we can expect that, according to the above remarks, the model fit is enhanced around this frequency. If $p_{0}=0$, corresponding to the classical basis $z^{-1}, z^{-2} \cdots$, one has $\chi(\omega)=e^{\omega}$, showing that $\chi(\omega)$ is maximum at the Nyquist frequency, and insignificant at low $\omega$; thus the model misfit at those frequencies plays a quasi negligible role in the minimization problem \eqref{e14}. This is the reason why classical PLR algorithms with basis $z^{-1}, z^{-2} \cdots$  generally exhibit important bias at low frequency and are absolutely not suited for fast sampled systems identification, hence the quite stringent rules regarding the sample period choice \cite{r13} that have been introduced for a long time. Likewise some specialists of PLR identification prefer to represent linear systems on Bode plots with a linear scale \cite{r4}, \cite{r13}.
 Figure \ref{fig1} displays the frequency distortion rate $\chi(\omega,p_0)$ corresponding to Laguerre bases  for many values of the Laguerre poles. One can observe the conservation principle of property \ref{prop6}.

\begin{figure}[H]
\begin{center}
\includegraphics[ width=3.4in, keepaspectratio ]{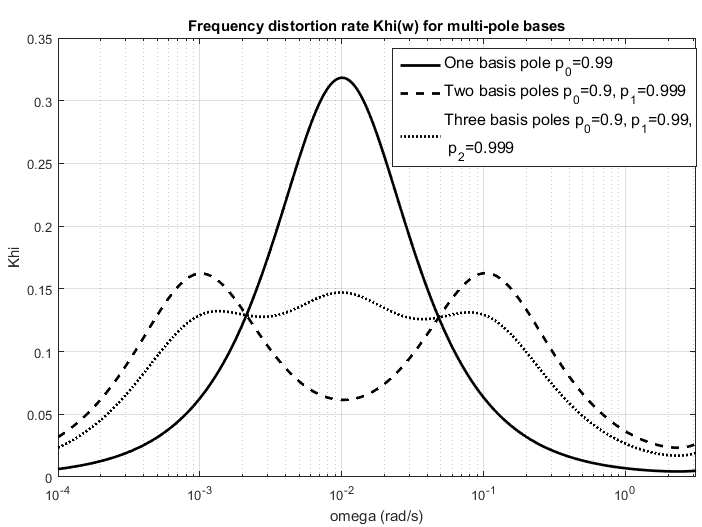}
\caption{Frequency distortion rate $\chi(\omega)$ for multi-poles bases}
\label{fig2}
\end{center}
\end{figure}

\noindent Figure \ref{fig2} shows three examples of  $\chi(\omega)$, corresponding to 1) one pole basis $p_{0}=0.99,$ 2) two poles basis with $p_{0}=0.9, p_{1}=0.999$, 3) three poles basis with $p_{0}=0.9, p_{1}=0.99, p_{2}=0.999$. This function $\chi$ provides a tool to assess qualitatively the effect of the basis pole on the model approximation in the frequency domain, and the said poles can therefore be considered as tuning parameters to specify, for a given experiment where to enhance  the model fit in the frequency domain.

\section{Simulation results}
\subsection{Identification of a reduced order system with a Laguerre basis}
\begin{figure}[H]
\begin{center}
\includegraphics[ width=3.5in, keepaspectratio ]{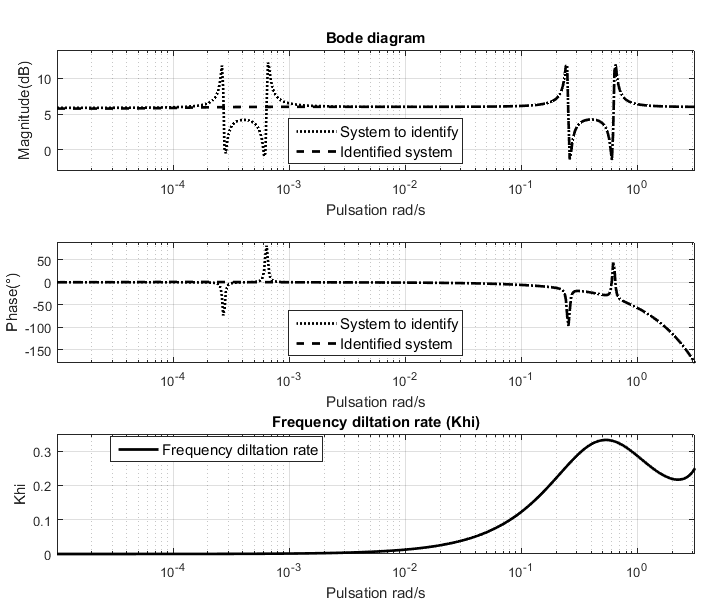}
\caption{Reduced order identification, recursive extended least squares, predictor with one pole basis, Laguerre pole $p_o=0.6.$}
\label{fig3}
\end{center}
\end{figure}

\begin{figure}[H]
\begin{center}
\includegraphics[ width=3.5in, keepaspectratio ]{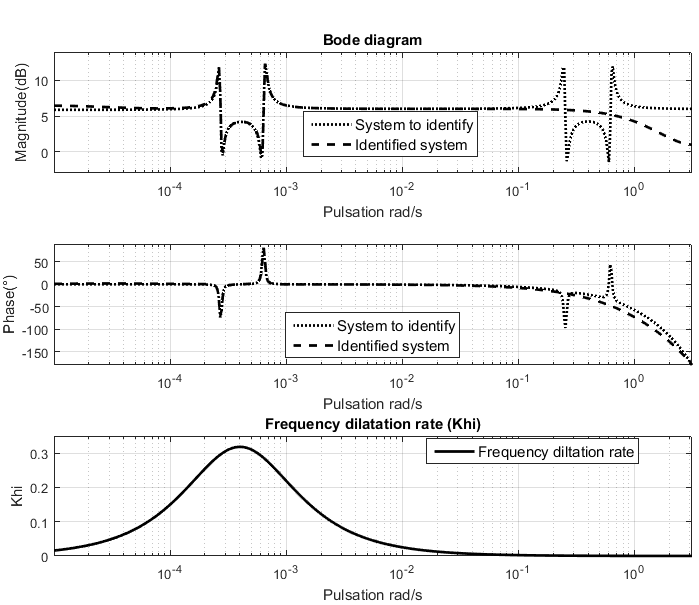}
\caption{Reduced order identification, recursive extended least squares, predictor with one pole basis, Laguerre pole $p_o=0.9996.$}
\label{fig4}
\end{center}
\end{figure}

In these simulations, the system to be identified consists of two clusters of two resonant and two antiresonant modes, separated from 3 decades, which corresponds clearly to a stiff system. The overall system has order equal to 9, and is disturbed by an white output noise (signal/noise ratio: 22 dB). We identify it by means of the H-ERLS algorithm (corresponding to an ARMAX model), and we choose a predictor parameterized with a one pole basis($\eta_p=1$), corresponding to a Laguerre basis. We look for a reduced order model $\eta_a=6$. The first simulation (figure \ref{fig3}) shows how the initial model is approximated if the Laguerre pole is chosen such that the frequency distortion rate maximum is near the high frequency modes ($p_o=0.6$), the system being excited by a PRBS (11 registers, length 2047 samples, no decimation). These high frequency modes are well captured, whereas the low frequency ones are sheerly ignored. On the contrary, if we set the Laguerre pole such that the frequency distortion rate is maximum at a frequency close to the low frequency modes ($p_0=0.9996$), we obtain a good model approximation at those frequencies as shown in figure \ref{fig4};  the system is excited by a 20 register PRBS, lenght $2^{20}-1$, without decimation, corresponding roughly to 44 periods of the lowest mode period (a lower noise level would allow for a lower test duration). This example shows that the frequency distortion rate function can be viewed as a tool allowing to appreciate the effect of the predictor parameterization on the bias distribution.

\subsection{Identification of a stiff system with a two poles basis}
\begin{figure}[H]
\begin{center}
\includegraphics[ width=3.5in, keepaspectratio ]{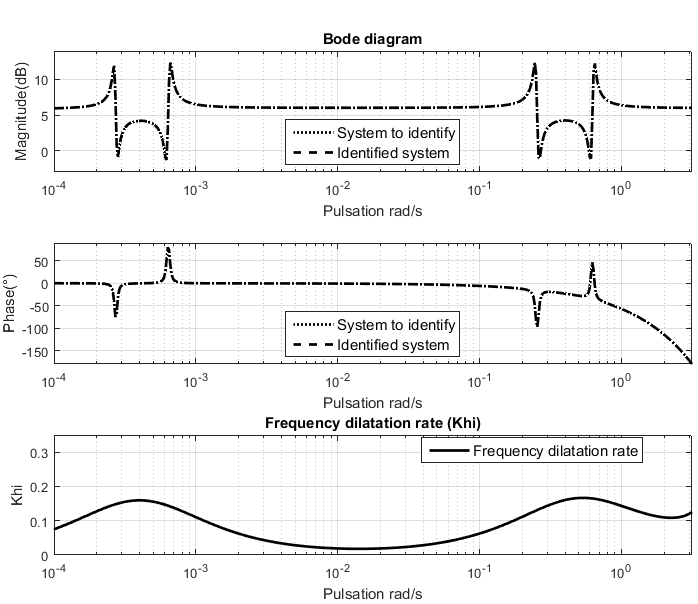}
\caption{Identification with a 10th order model, and a two poles basis  ($p_{0}=0.6, p_{1}=0.9996)$}
\label{fig5}
\end{center}
\end{figure}

Finally, we carry out an identification of the system aiming at capturing both low and frequency modes. This is made possible by selecting a 2 poles basis, and choosing a system order equal to 10 i.e. $\eta_p=2$, and $\eta_a=10$. The frequency distortion rate has now two maximal values, and we choose their frequencies in order to correspond roughly to those of low and high frequency modes clusters. The resulting identified model is displayed in Figure \ref{fig5}; this figure shows that a good fit has been obtained over all the spectrum, and that the model cannot be distinguished from the system to identify. The noise level is the same as in the previous subsection, and the excitation signal is a 20 register PRBS, (lenght $2^{20}-1$, without decimation).

\section{Conclusion}
In this paper, we have proposed a predictor parameterization of identification schemes belonging to the pseudo-linear regression class. This parameterization is established on an orthonormal transfer function basis, and it addresses Output Error, ARMAX and a generalization of ARX models. We have shown that the choice of the basis poles has a clear influence on the convergence conditions of recursive pseudo-linear algorithms. Moreover these poles modify the bias distribution of the estimated model. A method for assessing the basis poles effect on the bias distribution is presented; it is established on the analysis of the distortion between the classical frequency scale and the Hambo frequency one.  Successful  simulations of identifications performed on a stiff system show the interest of this analysis.

%


\bigskip

\begin{thebibliography}{99}

\bibitem[1]{r19} K.J. Astr{\"o}m, "On the choice of the sampling rates in parameter identification of time series"', Inf-Sci, p.273-278 (1969).

\bibitem[2]{r22} H. Garnier, R.R. Bitmead, R.A. de callafon, "`Direct continuous-time model identification of high-powered night-emitting diodes rapidly sampled thermal step response", Proc. of the 19thIFAC world congress, cape town 2014.

\bibitem [3] {r3} P.S.C Heuberger, P.M.J Van den Hof, O.H. Bosgra, "A generalized orthonormal basis for linear dynamical systems"', IEEE, Trans. on automatic control, vol. 40, pp. 451-465, 1995.

\bibitem [4] {r1} P.S.C Heuberger, P.M.J Van den Hof, B. Wahlberg, \textit{Modelling and identification with rational orthogonal basis functions}, Springer Verlag, 2005.

\bibitem [5] {r2} P.S.C Heuberger, T.J. de Hoog, P.M.J van den Hof, B. Wahlberg, "Orthonormal basis functions in time and frequency domain: Hambo transform theory"', SIAM, J. Control and opt., vol42(4), pp. 1347-1373, 2003.

\bibitem[6]{r16} I.D. Landau, "Unbiased recursive identification using model reference adaptive techniques", IEEE transactions on automatic control,vol 21(2), pp. 194-202 (1976). 

\bibitem[7]{r17} I.D. Landau, A.Karimi, "An output error recursive algorithm for unbiased identification in closed-loop"'. Automatica 33(5), pp. 933-938 (1997).

\bibitem[8]{r18} I.D. Landau, A. Karimi, "`A recursive algorithm for ARMAX model identification in closed-loop"'. IEEE. Trans on automatic control, vol 44(4), pp.840-843.

\bibitem [9] {r4} I.D. Landau, R. Lozano, M. M'Saad, A. Karimi, \textit{Adaptive control, second edition}, Springer Verlag, 2011.

\bibitem [10]{r13} I.D. Landau, G. Zito, \textit{Digital control systems}, Springer 2006.

\bibitem[11]{r20} Ph. de Larminat, \textit{Automatique appliqu\'ee}, Herm\`es, (in French), 2009.

\bibitem [12]{r11} L. Ljung, \textit{System identification, theory for the user (second
edition), }Upper Saddle River, Prentice Hall, 1999. 

\bibitem [13]{r6} L.Ljung, T. S{\"o}derstr{\"o}m, \textit{Theory and practice of recursive identification, }Cambridge, MIT\ Press, 1983. 

\bibitem[14]{r15} V. Panuska, "A stochastic approximation method for identification of linear systems using adaptive filtering", in Joint Automatic control conference, pp. 1014-1021, AnnArbor, Mic (1968).

\bibitem [15]{r12} F.Shipp, L. Gianone, J. Bokor, Z. Szabo, "`Identification in generalized orthogonal basis- a frequency domain approach. preprints of the 13th IFAC World congress, vol(1), p.387-392, San Francisco, CA, 1996, Elsevier.

\bibitem[16]{r23} T. O. e Silva, "`Laguerre filters-An introduction", Revista do Detua, vol.1, no. 3, pp. 237-248, jan. 1995.


\bibitem [17]{r9} B.Vau, H. Bourl\`es, "Some remarks on the bias distribution analysis of discrete-time identifcation algorithms based on pseudo-linear regressions", Systems and Control Letters, vol. 119, pp.46-51, 2018.

\bibitem [19]{r7} B.Vau, H. Bourl\`es, "Laguerre based predictors in discrete-time recursive algorithms: A solution for open-loop identification under oversampling", 20th IFAC world congress, Toulouse, France, 2017.

\bibitem[20]{r21} B. Wahlberg, "System identification using Laguerre models", IEEE Trans.on automatic control, vol. 38(9), pp. 1371-1383, (1991).

\bibitem [21]{r10} B.Wahlberg, L.Ljung, "Design variables
for bias distribution in transfer function estimation'', IEEE transactions on automatic control, vol. 31(2), pp. 134-144, 1986. 

\bibitem[22]{r14} P.C. Young, "The use of linear regression and related procedures for the identification of dynamic processes" Proc. of the 7th IEEE Symposium on adaptive process, pp. 501-505, San Antonioa, Texas (1968).
























\end{thebibliography}
\end{document}